\documentclass[fleqn,twoside]{article}
\usepackage{espcrc2}
\usepackage{url}
\usepackage[numbers]{natbib}
\usepackage{graphicx}
\newcommand{\Xmax}{\ensuremath{X_{\rm max}}}
\newcommand{\Nmax}{\ensuremath{N_{\rm max}}}

\title{Observation of the GZK Cutoff Using the HiRes Detector}

\author{D.~R.~Bergman
  \address[Rutgers]{Rutgers, 
    The State University of New Jersey  \\
    Department of Physics and Astronomy \\
    Piscataway, New Jersey, USA 08854}%
  \thanks{e-mail:  bergman@physics.rutgers.edu}
  presented on behalf of the High Resolution 
    Fly's Eye Collaboration}

\begin{document}

\begin{abstract}
  The High Resolution Fly's Eye (HiRes) experiment has observed the
  GZK cutoff.  HiRes observes two features in the ultra-high energy
  cosmic ray (UHECR) flux spectrum: the Ankle at an energy of
  $4\times10^{18}$ eV and a high energy suppression at
  $6\times10^{19}$ eV.  The later feature is at exactly the right
  energy for the GZK cutoff according to the $E_{1/2}$ criterion.
  HiRes cannot claim to observe a third feature at lower energies, the
  Second Knee.  The HiRes monocular spectra are presented, along with
  data demonstrating our control and understanding of systematic
  uncertainties affecting the energy and flux measurements.
  \vspace{1pc}
\end{abstract}

\maketitle

HiRes has observed the GZK Cutoff.  The GZK cutoff was predicted 40
years ago, Greisen~\cite{Greisen-1966-PRL-16-748} and Zatsepin and
Kuzmin~\cite{Zatsepin-1966-JETPL-4-78} predicted the existence of a
sharp reduction in the flux of ultra-high energy cosmic rays (UHECR's)
at energies above about $6\times10^{19}$ due to photopion production
interactions between UHECR protons and the cosmic microwave background
radiation (CMBR).  

Since the prediction of the GZK cutoff was made, a number of
experiments claim to have observed events with measured energies above
$1\times10^{20}$ eV.  Volcano Ranch\cite{Linsley-1963-PRL-10-146}
observed one event in 1963, even before the GZK Cutoff was proposed.
However, the knowledge of the expected lateral distribution function
to use in estimating the total number of shower particle was lacking
at that time.  SUGAR claimed three events\cite{Winn-1986-JPG-12-653}
and Haverah Park observed four\cite{Lawrence-1991-JPG-17-733}.  SUGAR
used only muon detectors, however, and had problems with after pulsing
in its PMT's.  One should ``be cautious about taking the energies
ascribed to the Sydney events \dots as providing definitive evidence
against a cutoff in the cosmic ray energy
spectrum\cite{Nagano-Watson-2000-RMP-72-689}.''  The Haverah Park data
was later reanalyzed\cite{Ave-2003-APP-19-47} using Corsika to
establish the relation between $\rho_{600}$ and the energy; all the
high energy events were reanalyzed to lower energies, below the GZK
cutoff.  More recently, both the Yakutsk
Array\cite{Egorova-2004-NPBps-136-3} and
AGASA\cite{Takeda-2003-ICRC-28-381} have claimed to events above
$10^{20}$ eV, but Yakutsk claims that it's events are consistent with
what is expected from the GZK process while AGASA had claimed
otherwise, but is now reconsidering their energy
scale\cite{Teshima-2006-CRIS}.

The point of going through this short history is to note that it is
\emph{hard} to measure the energy well with a surface array (which all
of the above are), because one is never looking at the \emph{bulk} of
the shower.  This makes the energy determination subject to the poorly
understood systematics of shower modeling, and it is necessary to
measure the energy well to observe a break in such a steeply falling
spectrum.  It is easier, on the other hand, to measure the energy
well with a fluorescence detector, such as HiRes.  Because a
fluorescence detector does observe the bulk of the shower, one is left
only with systematic uncertainties which are easier to understand and
control.

Two such systematic uncertainties are number of photons emitted by the
shower as it passes through the atmosphere, the fluorescence yield,
and the clarity of the atmosphere through which one observes those
emitted photons.  The fluorescence yield has been well studied in the
lab, and the differences between three recent measurements is only at
the 6\% level
\cite{Nagano-2003-APP-20-293,Kakimoto-1996-NIMA-372-572,Belz-2006-APP-25-129}.
The atmospheric uncertainty, on the other hand, can be controlled by
going to a site with a clean stable atmosphere, like Dugway, Utah, and
then by measuring the scattering properties of the air with lasers as
we have done in HiRes\cite{Abbasi-2006-APP-25-74}.

A third systematic uncertainty for a fluorescence detector is the size
of the aperture as a function of energy.  Traditionally, ground arrays
were thought to have constant aperture above the energy where the
array becomes 100\% efficient.  In fact, given energy and core
location resolutions this is not quite true.  Fluorescence detectors,
however, have a manifestly changing aperture because higher energy
showers are brighter and can be seen farther away.  The problem then
becomes one of determining the size of the aperture at a given energy
by a Monte Carlo simulation.  However, this is not a case where we
have replaced one poorly understood simulation, extensive air shower
development, which is the bane of ground array experiments, with
another poorly understood simulation.  The aperture calculation for a
fluorescence detector uses only well understood physics and, more
importantly, is amenable to detailed \emph{verification} through the
technique of Data/MC Comparisons.  It is just these detailed
comparisons that comprise the bulk of this talk and give us the great
confidence we have in our aperture calculation.

I will begin this discussion of the aperture calculation by noting
that the aperture is, in accelerator physics parlance, the acceptance
of the detector, and that acceptances are routinely calculated by MC
simulation:
\begin{eqnarray}
{\rm Ap}(E') &=& \frac{N_R(E')}{N_T(E)}
  {\rm Area}_T\Omega_T \\
            &=& {\rm Acpt(E')}{\rm Area}_T\Omega_T 
\end{eqnarray}
where $E'$ is the reconstructed energy, while $E$ is the thrown
energy.  $N_T$ and $N_R$ are the numbers of events generated
(``thrown'') and reconstructed, respectively.  For the calculation to
be accurate, one must put the right distribution of showers in the
right places (to get the area), at the right angles (to the right
solid angle), and with the right amount of light (to get the right
energy scale).  What is more, one must put in the right fluctuations
in signal for the difference between reconstructed energy ($E'$) and
thrown energy ($E$) to be significant.  Fortunately, these
distributions are easy to check in Data/MC Comparisons.

\begin{figure}[t]
  \includegraphics[width=1.00\columnwidth]{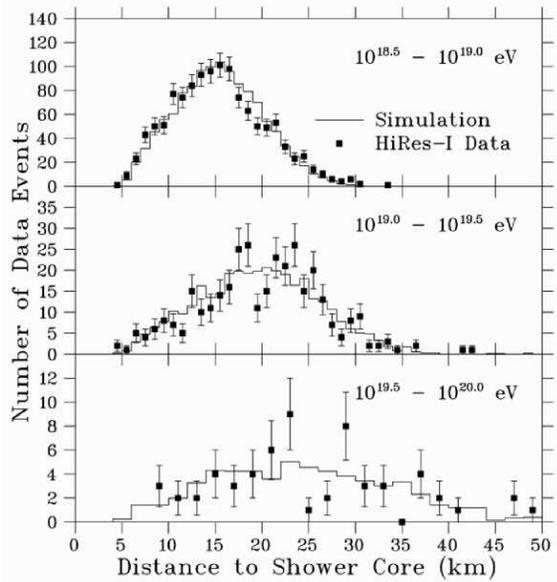}
  \caption{A Data/MC Comparison of the distance to the impact point from
    HiRes-I.  The three panels show events in the different energy
    ranges as labeled.  Data is shown as points with error bars, the
    MC as a histogram.}
  \label{hr1-rimp}
\end{figure}

\begin{figure}[ht]
  \includegraphics[width=1.00\columnwidth]{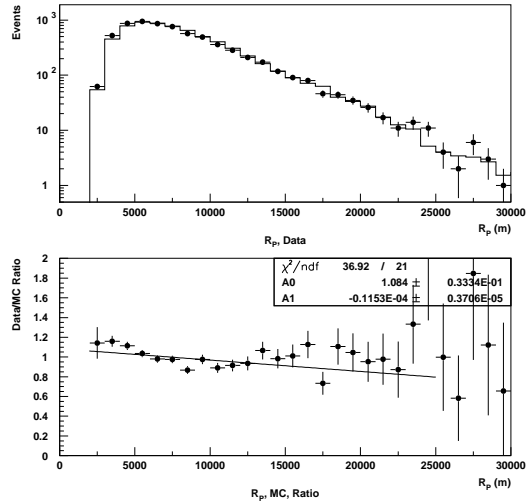}
  \caption{A Data/MC Comparison of the impact parameter, $R_P$, from
    HiRes-II.  The top panel shows the data as points with error bars,
    the MC as a histogram.  The MC has about five times the statistics
    as the data.  The bottom panel shows the ratio of data to MC, with
    a linear fit over bins with more than 10 data events.}
  \label{hr2-rp}
\end{figure}

\begin{figure}[ht]
  \includegraphics[width=1.00\columnwidth]{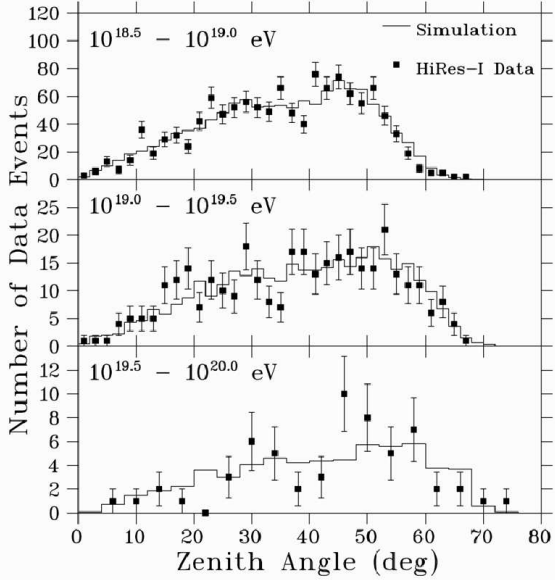}
  \caption{A Data/MC Comparison of the zenith angle in HiRes-I.  The
    three panels show events in the different energy ranges as
    labeled.  Data is shown as points with error bars, the MC as a
    histogram.}
  \label{hr1-zenith}
\end{figure}

\begin{figure}[ht]
  \includegraphics[width=1.00\columnwidth]{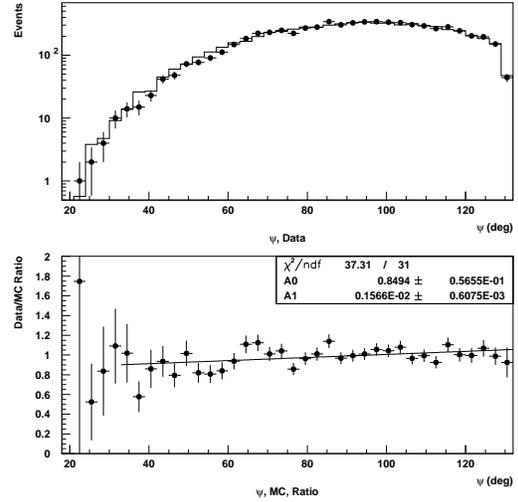}
  \caption{A Data/MC Comparison of the shower-detector plane angle,
    $\psi$, from HiRes-II.  The top panel shows the data as points
    with error bars, the MC as a histogram.  The MC has about five
    times the statistics as the data.  The bottom panel shows the
    ratio of data to MC, with a linear fit over bins with more than 10
    data events.}
  \label{hr2-psi}
\end{figure}

\begin{figure}[ht]
  \includegraphics[width=1.00\columnwidth]{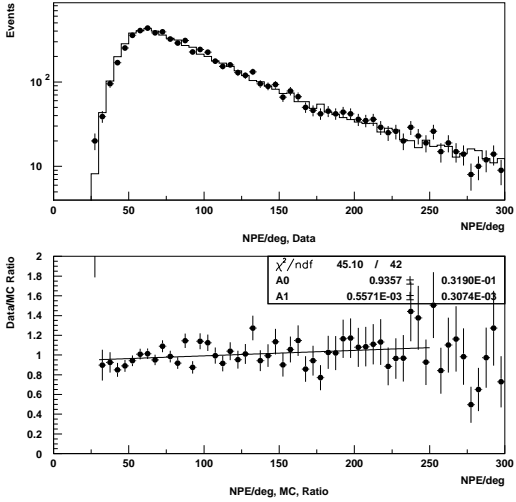}
  \caption{A Data/MC Comparison of the event brightness, total signal
    divided by track length, for HiRes-II.  The top panel shows the
    data as points with error bars, the MC as a histogram.  The MC has
    about five times the statistics as the data.  The bottom panel
    shows the ratio of data to MC, with a linear fit over bins with
    more than 10 data events.}
  \label{hr2-chi2}
\end{figure}

\begin{figure}[ht]
  \includegraphics[width=1.00\columnwidth]{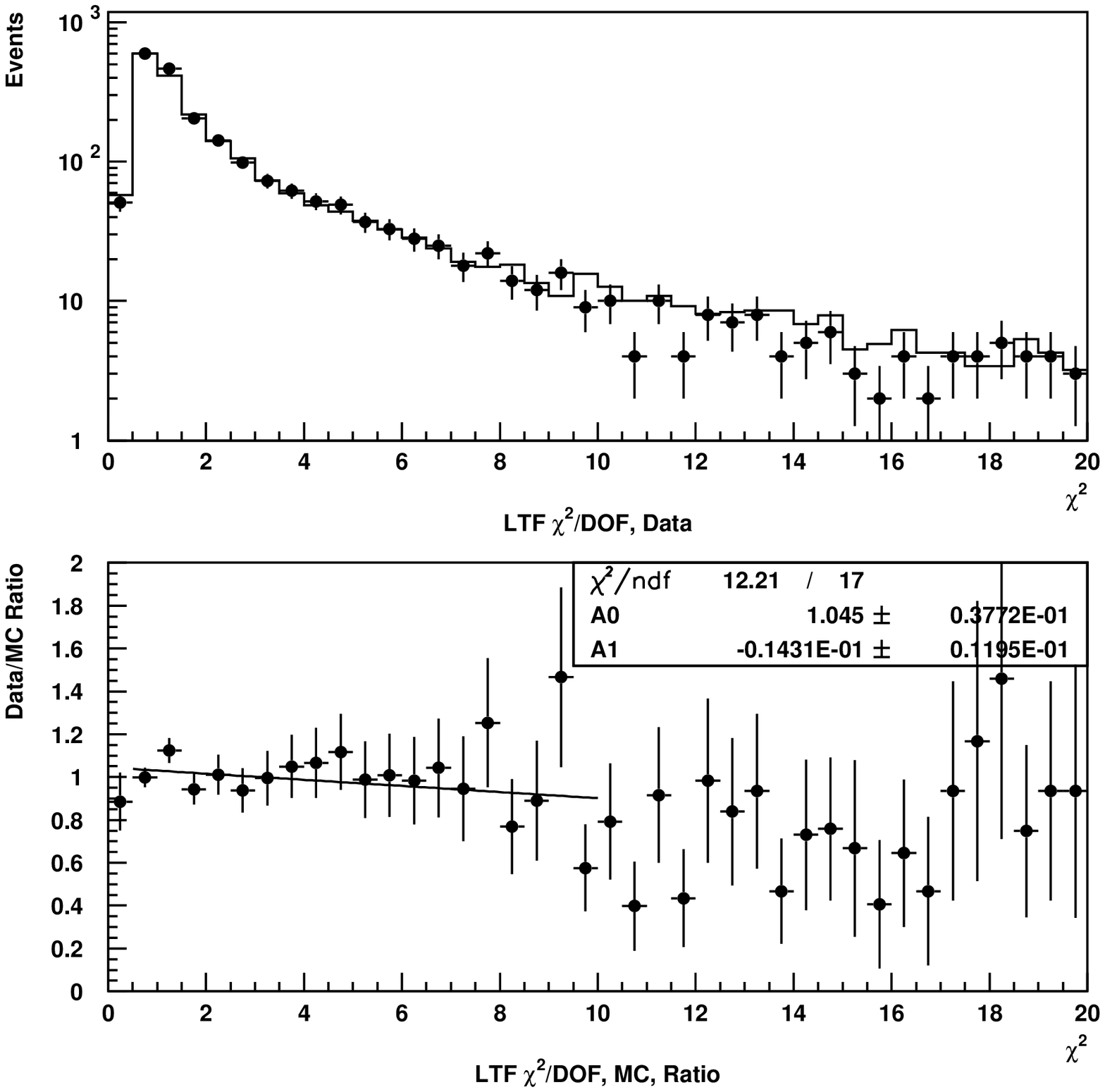}
  \caption{A Data/MC Comparison of the distribution of $\chi^2/{rm
      DOF}$ for a linear fit of tube time vs angle, for HiRes-II.  The
    top panel shows the data as points with error bars, the MC as a
    histogram.  The MC has about five times the statistics as the
    data.  The bottom panel shows the ratio of data to MC, with a
    linear fit over bins with more than 10 data events.}
  \label{hr2-npedeg}
\end{figure}

The first comparisons are for the distances of the showers from both
HiRes-I and HiRes-II (events in the right \emph{places}).  These are
shown in Figure~\ref{hr1-rimp} for the distance to the impact point in
HiRes-I in three energy bins, and in Figure~\ref{hr2-rp} for the
impact parameter for showers from HiRes-II.  The second set of
comparisons shows the distributions of measured angles (events at the
right \emph{angles}): the zenith angle for HiRes-I events in
Figure~\ref{hr1-zenith}, the angle between the shower and the ground
in the shower-detector plane, $\psi$, for HiRes-II events in
Figure~\ref{hr2-psi}.  A third comparison, in Figure~\ref{hr2-npedeg}
shows the brightness distribution for events at HiRes-II (events with
the right amount of light).  The final comparison, in
Figure~\ref{hr2-chi2}, shows the distribution of $\chi^2$ for a fit
of the times of tubes at HiRes-II to a linear function of time vs.
angle (for resolution).  Since the time .vs angle relation should be
curved, the $\chi^2$ distribution has a large tail.

\begin{figure}[ht]
  \includegraphics[width=1.00\columnwidth]{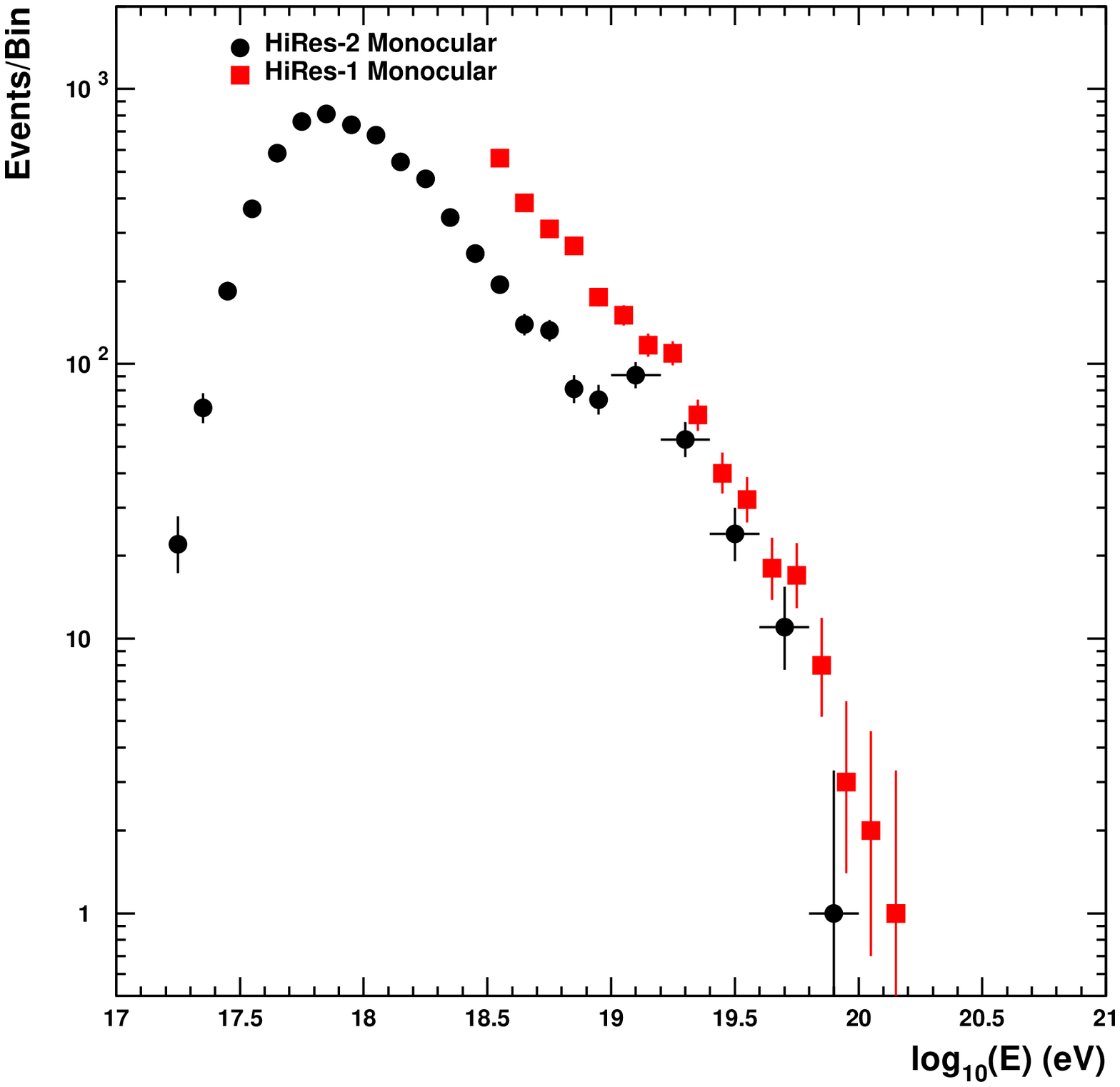}
  \caption{The number of events collected in each energy bin in
    monocular mode by HiRes-I and by HiRes-II.  The jump in HiRes-II
    statistics at $10^{19}$ eV is due to the bins being twice as big
    in $\log E$.}
  \label{hr12-events}
\end{figure}

\begin{figure}[ht]
  \includegraphics[width=1.00\columnwidth]{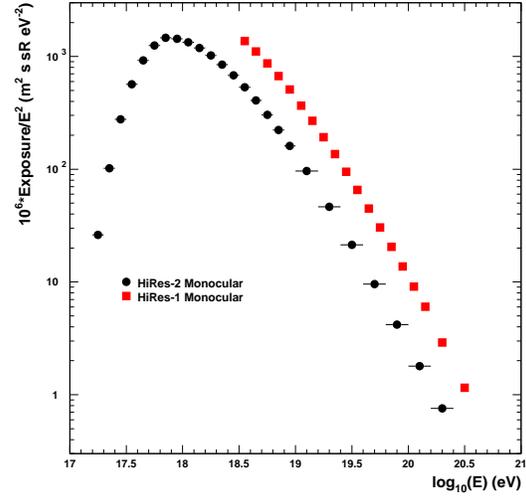}
  \caption{The calculated exposure of HiRes-1 and HiRes-II in
    monocular mode.  The exposure has been divided by $E^2$ to make
    direct comparisons to the event distribution easier.}
  \label{hr12-expos}
\end{figure}

\begin{figure}[ht]
  \includegraphics[width=1.00\columnwidth]{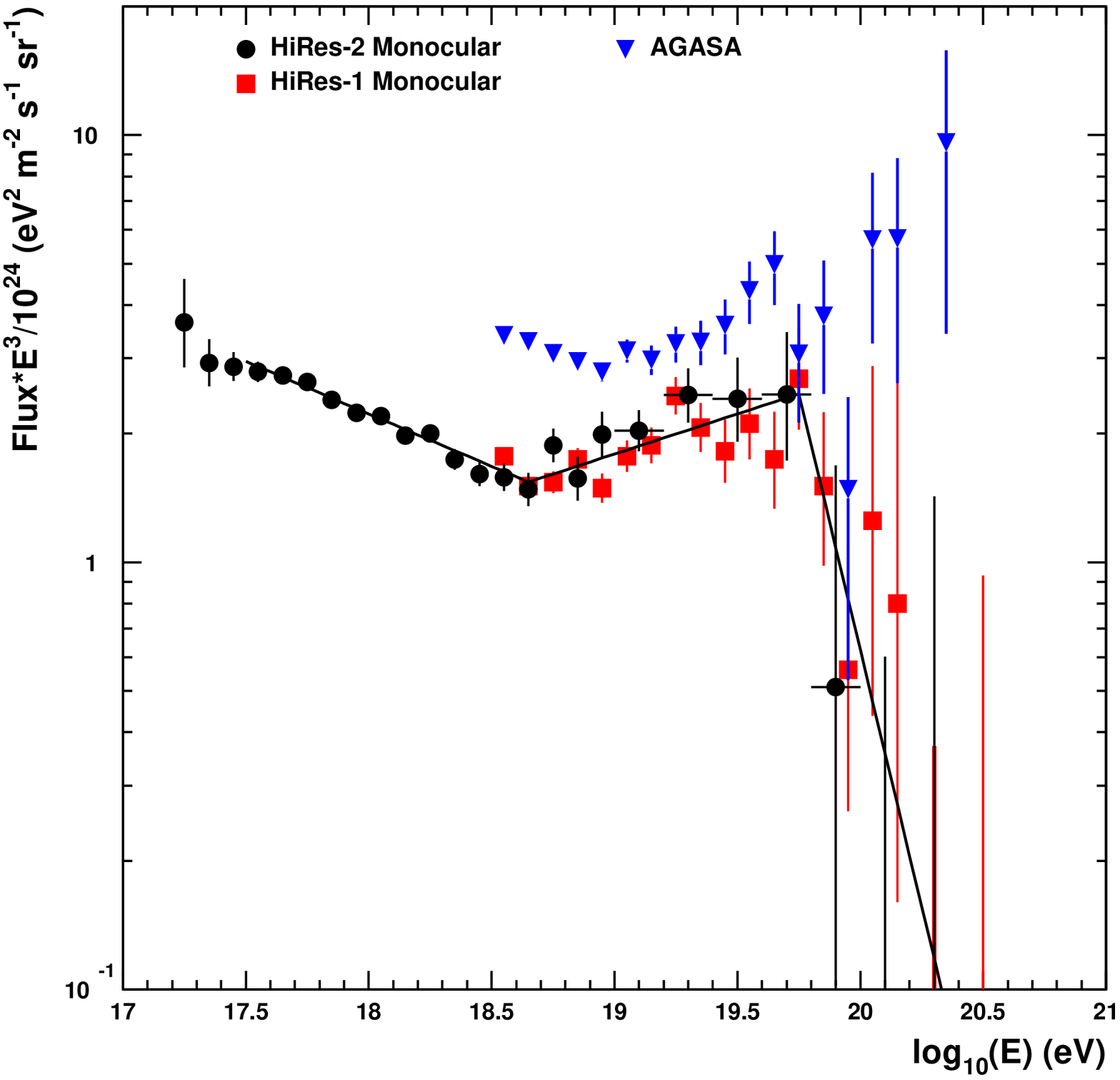}
  \caption{The measured UHECR spectra from HiRes-I and HiRes-II in
    monocular mode.  Also shown is a recent AGASA spectrum for
    comparison.  The HiRes-I and HiRes-II spectra have been fit to a
    broken power law spectrum with two break points.  The result of
    this fit is also shown.}
  \label{hr12ag-trilinfit}
\end{figure}

Having built confidence in our aperture calculation, we move on to the
data measured by HiRes in monocular mode.  This is shown in
Figure~\ref{hr12-events}.  It is \emph{this} distribution which we
will be fitting later to determine the significance of our observation
of the GZK Cutoff.  The calculated exposure (aperture times time) is
shown in Figure~\ref{hr12-expos}.  Dividing the event distribution by
the exposure gives the spectrum, which is shown in
Figure~\ref{hr12ag-trilinfit}.

We performed a number of systematic cross checks on these data to
ensure their accuracy.  First we limited the distance to showers to 10
and 15 km.  We saw no systematic change in the spectra in these cases,
only variations due to reduced statistics.  What is more, the Ankle
feature is visible in the event distribution in the 10 km case.
Second, we made a spectrum from the events which are in both the
HiRes-I and HiRes-II samples, using the precision shower geometry
measurement available in this case.  All the other details of the
analysis are the same as in the monocular analyses, and the spectrum
agrees with our monocular spectra.  This shows that the energy
resolution in monocular mode is good enough not to distort the
spectrum from the stereo measurement beyond the statistical power of
the data we have collected.  This was \emph{not} true of Fly's Eye.
We also tested the aperture at 35 km, by reconstructing vertical laser
shots at this distance.  A variety of laser intensities were used,
which we linked to cosmic ray energies based on the brightness of the
tracks.  Our detector becomes fully efficient for vertical showers at
35 km for showers with energies above 60 EeV.

\begin{figure}[ht]
  \includegraphics[width=1.00\columnwidth]{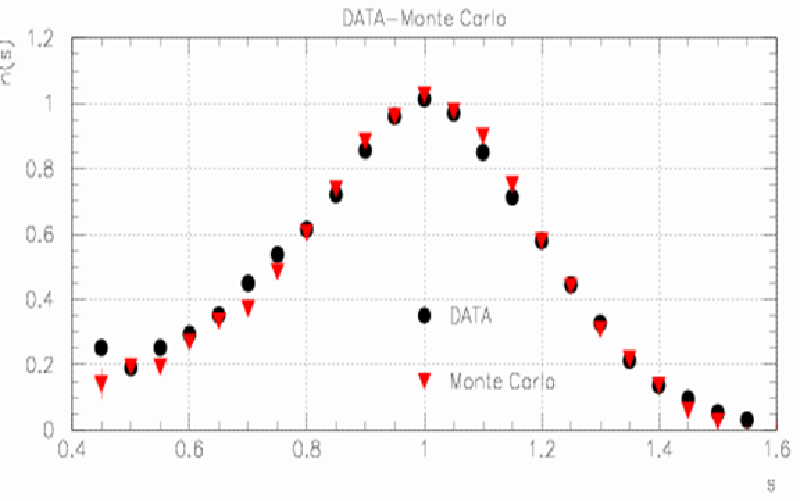}
  \caption{A Data/MC comparison of the average profile for showers in
    HiRes-II.}
  \label{datamc-avgprof}
\end{figure}

Because we use a fit the Gaisser-Hillas profile to determine the
shower energy, we compared the average shower profile in data to that
calculated in MC using the Gaisser-Hillas profile.  In each case the
showers are normalized by \Xmax\ (using the age parameter) and \Nmax.
Then the actual shower measurements are averaged in bins of the age
parameter.  The comparison is shown in Figure~\ref{datamc-avgprof}.
This shows that the Gaisser-Hillas profile is appropriate to use for
measuring shower energies.

To check the systematic uncertainty from shower modeling, we
calculated the aperture using both QGSJet and
Sibyll\cite{Abbasi-2006-astro-ph-0607094}.  It is important to note
that we adjust the proton-to-iron ratio in each case to give identical
mean Xmax values as a function of energy.  With this requirement, we
calculate identical apertures for both models within the MC statistics
(5$\times$ the data) available.

We cannot claim to observe the Second Knee in the HiRes-II spectrum.
The data statistics are limited and the aperture has a larger
uncertainty than at energies above $10^{18}$ eV due to composition
uncertainties and their effect on the
aperture\cite{Abbasi-2006-astro-ph-0607094}.  Because of this, we will
only be fitting the spectra above $10^{17.5}$ eV.

We now move on to broken power law fits of the two HiRes monocular
spectra.  We will be fitting the spectra using a binned maximum
likelihood method which uses the number of observed events in a bin
and the number of events expected.  This will allow us to use empty
bins in the data where there is a significant number expected from the
power law.  It also allows us use both data samples independently.  A
correction will be made at the end for events seen by both detectors.
The number of expected events is given by the broken power law flux
times the measured exposure.  We will begin with no break points and
add floating break points until there is no improvement in the
$\chi^2$.

A fit of both datasets to a simple power law gives a $\chi^2$ of 162
for 39 degrees of freedom, with a spectral index $\gamma$ of
$3.13\pm0.01$.  This is clearly not a good fit.  Adding one floating
break point gives a much better fit, and the break point finds the
position of the Ankle.  The $\chi^2$ is now 68 for 37 degrees of
freedom, with the break point at $4.3\pm0.5$ EeV, a spectral slope
below the Ankle of $\gamma_1 = 3.24\pm0.02$ and a spectral slope above
the Ankle of $\gamma_2 = 2.89\pm0.03$.  Adding a third break point
reduces the $\chi^2$ by 33 to 34.7 for 35 degrees of freedom.  The
breakpoints are at 4.5 EeV and 56 EeV, while the three spectral
indices are $3.24\pm0.02$, $2.81\pm0.03$ and $5.4\pm0.7$.  This fit is
shown in Figure~\ref{hr12ag-trilinfit}.  Adding a third breakpoint
does not significantly reduce the $\chi^2$.

We calculate the significance of the break at 56 EeV by comparing the
number of events expected above $10^{19.8}$ eV if the spectral index
continued at 2.81 to the actual number of events observed.  In this
case we expect 51.1 events but observe only 15.  The Poisson
probability of observing 15 or fewer events when expecting 51.1 is
$2.9\times10^{-9}$.  In actuality, we have counted one event observed
in stereo twice in our 15 observed events and have also counted
expected events from overlapping exposure from the two sites.  When
each of these is taken into account we get a revised expectation of
44.9 events while observing 14.  The Poisson probability in this case
is $7\times10^{-8}$.  For comparison the fraction of the area in one
tail of a Gaussian beyond $5\sigma$ is $3\times10^{-7}$ and beyond
$6\sigma$ is $1\times10^{-9}$.  Thus we claim a significance of over
$5\sigma$ for our measurement of a high energy break point in the
spectrum.

\begin{figure}[ht]
  \includegraphics[width=1.00\columnwidth]{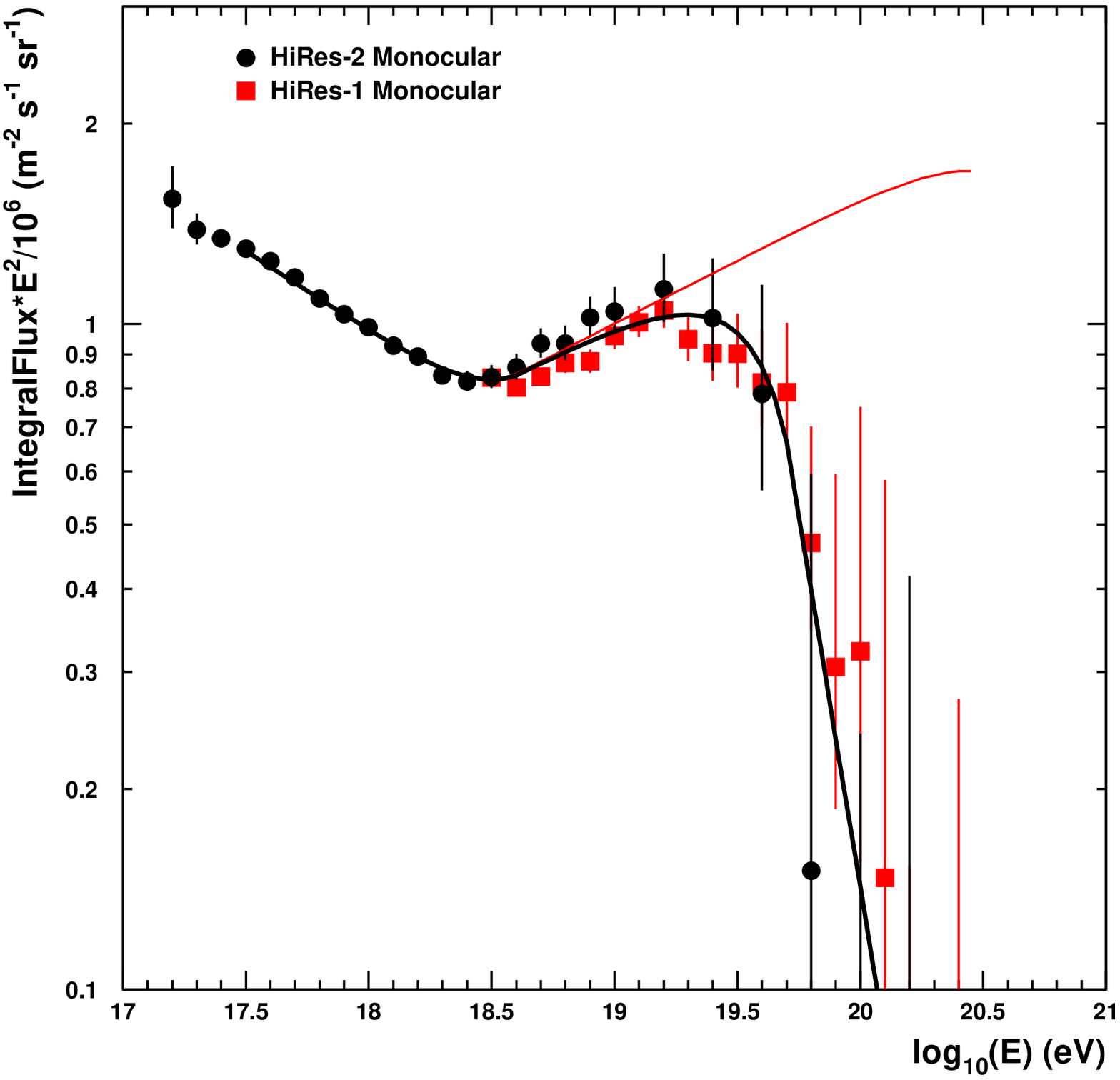}
  \caption{The integral spectra measured by HiRes-I and HiRes-II.
    Also shown are the integrated broken power law fits to the
    differential spectra.  The black line shows the fit with two break
    points; the red line shows the same fit in which the high energy
    break point has been removed with no change in the other
    parameters.}
  \label{hr12-ie2-tlfx}
\end{figure}

\begin{figure}[ht]
  \includegraphics[width=1.00\columnwidth]{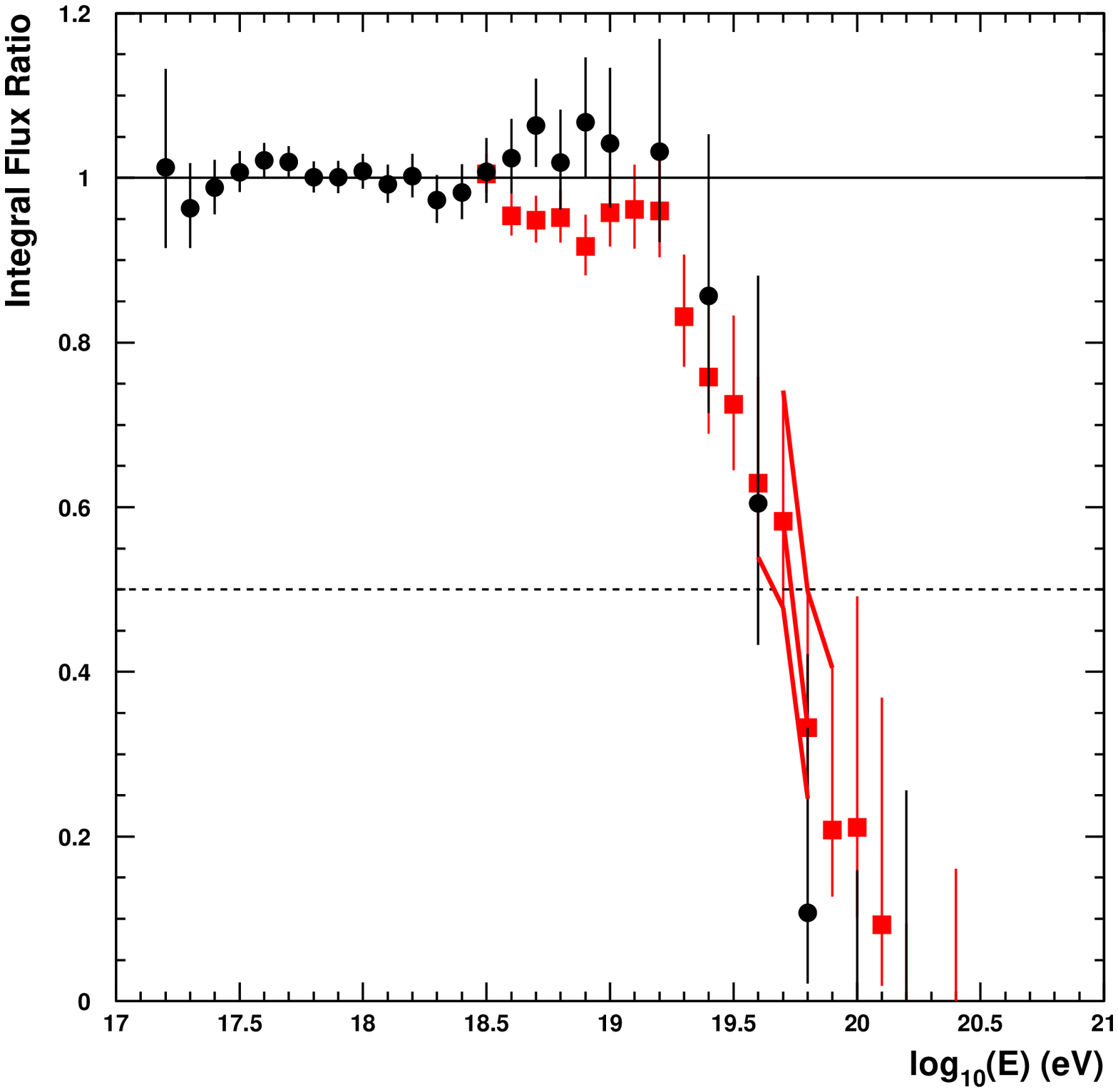}
  \caption{The ratio of the HiRes-I and HiRes-II integral spectra to
    the expected integral flux in the absence of a GZK cutoff.  Only
    the HiRes-I values are used to make an estimate of $E_{1/2}$ by
    interpolating between the central value and $1\sigma$ uncertainty
    limits. }
  \label{hr12-ie2-ratio}
\end{figure}

One way to compare the energy of this breakpoint to that expected of
the GZK Cutoff is to use the $E_{1/2}$ method of
Berezinsky\cite{Berezinsky-Grigoreva-1988-AA-199-1}.  $E_{1/2}$ is the
energy at which the integral spectrum falls to half of what would be
expected in the absence of the feature.  This HiRes monocular spectra
are displayed as integral spectra in Figure~\ref{hr12-ie2-tlfx} along
with the two break point fit mentioned above and the extension of this
fit used to measure the significance of the high energy break point.
Taking the ratio of the measured points to the extended fit allows us
to measure $E_{1/2}$.  This is shown in Figure~\ref{hr12-ie2-ratio}.
We find $\log E_{1/2} = 19.73\pm0.07$.  The prediction Berezinsky {\it
  et al}\cite{Berezinsky-Grigoreva-1988-AA-199-1} for the GZK effect
for a wide range of spectral indices is $\log E_{1/2} = 19.72$.  In
this way we have linked the energy of our measurement of a high energy
break point to the GZK Cutoff.

In conclusion, HiRes has made a measurement of the UHECR spectrum
using its two detectors in monocular mode.  The aperture used for this
measurement is well understood as verified by Data/MC comparisons.  We
observe a high energy break in the spectrum at the energy to be the
GZK Cutoff and with a significance of over $5\sigma$.  Thus we claim an
observation of the GZK Cutoff.

This work is supported by US NSF grants PHY-9321949, PHY-9322298,
PHY-9904048, PHY-9974537, PHY-0098826, PHY-0140688, PHY-0245428,
PHY-0305516, PHY-030098, and by the DOE grant FG03-92ER40732.We
gratefully acknowledge the contributions from the technical staffs of
our home institutions. The cooperation of Colonels E. Fischer and G.
Harter, the US Army, and the Dugway Proving Ground staff is greatly
appreciated.

\bibliography{UHECR-Spectrum}

\end{document}